\documentstyle[prc,preprint,tighten,aps]{revtex}
\def\braket<#1>{\langle{#1}\rangle}
\def\ketbra|#1><#2|{|{#1}\rangle\langle{#2}|}
\def\ket|#1>{|{#1}\rangle}
\def\vec#1{{\bbox{#1}}}
\def\tfrac#1/#2{{\textstyle\frac{#1}{#2}}}
\def\Li#1{\mbox{$^{#1}$Li}}
\def\Lipol#1{\mbox{$^{#1}\roarrow{\rm Li}$}}

\begin{document}

\draft

\title{Nucleon polarization in three-body models of polarized \bbox{^6}Li}

\author{N.W. Schellingerhout\cite{email} and L.P. Kok}
\address{Institute for Theoretical Physics, University of Groningen,
Nijenborgh 4, NL-9747~AG~~Groningen, The Netherlands}
\author{S.A. Coon}
\address{Department of Physics, New Mexico State University,
Las Cruces, New Mexico, 88003}
\author{R.M. Adam\cite{RobAddress}}
\address{Department of Physics and Astronomy, Vrije Universiteit,
De Boelelaan 1081, NL-1081~HV~~Amsterdam, The Netherlands}

\date{\today}

\maketitle

\begin{abstract}

\noindent
Just as $^3\roarrow{\rm He}$ can be approximately characterized as a
polarized neutron target, polarized \Li6D has been advocated as a
good {\em isoscalar} nuclear target for the extraction of the
polarized gluon content of the nucleon. The original argument rests
upon a presumed ``alpha + deuteron'' picture of \Li6, with the
polarization of the nucleus carried by the polarization of the
deuteron. We have calculated the polarization of the constituents of
\Li6 as a three-body bound state of $\alpha + n + p$ interacting with
local potentials fitted to the scattering data. It is necessary to
include partial waves up to $j=17/2$ (75 channels, or, when including
the $T=1$ state, 150 channels) in the Faddeev equations before the
energy eigenvalue converges. The longitudinal formfactors are then
described well by the wave function. Various combinations of $\alpha$N
and NN strong and Coulomb potentials yield a straight line in the
charge radius {\em vs.} energy plane which, unlike those of previous
calculations, passes through the experimental datum. We find for all
cases a polarization of the valence neutron in excess of 90\%. This
may make polarized \Li6D an attractive target for many nuclear physics
purposes, since its neutrons are effectively 45\% polarized.

\end{abstract}



\narrowtext

\section{Introduction}

Targets of polarized nuclei are rather rare, so the recently acquired
ability to produce a large solid target of polarized \Li6D
\cite{exppol} has aroused much interest in the medium-energy and
high-energy communities. In particular, this target has been suggested
for measurements of direct photon production at Fermilab
\cite{fermilab}, and at the UNK proton accelerator under construction
at Serpukhov. The cross section for direct photon production is
dominated by the ``Compton-like'' subprocess ``quark + gluon
$\rightarrow$ gamma + quark''. When both beam and target nucleons are
longitudinally polarized, asymmetries in this process can be used to
determine the polarized gluon distribution in a nucleon. Knowledge of
this spin-dependent (polarized) gluon distribution is needed to
extract the true spin-dependent quark and antiquark distributions from
the deep-inelastic lepton-scattering data \cite{EMC,HERMES}, and
establish the relationship of the polarization of a nucleon to that of
its quark and gluon constituents. For that particular high-energy
experiment, the most sensitive experimental method would be to use a
polarized isoscalar nucleon target, {\it e.g.}, one made up of equal
numbers of polarized protons and neutrons \cite{berger}. Polarized
\Li6D has been advocated as a good isoscalar nucleon target since ``To
the extent that \Lipol6 can be viewed as He + $\roarrow{\rm D}$, as
much as one-half of the nucleons are polarized in such a material''
\cite{fermilab}. (This is important from an experimental viewpoint
because, by contrast, conventional polarized-proton-target materials
contain less than 20\% free hydrogen by weight.) One can ask the
question ``To what extent, indeed, is this true?''  One of the goals
of this paper is to provide a theoretical analysis and prediction of
the polarization of the constituents of \Lipol6 in a dynamical
three-body model of \Li6. In this picture \Li6 is a bound system of an
alpha particle, a neutron, and a proton interacting with potentials
which parametrize the free space forces between the three
``elementary'' particles. We find that the optimistic polarization
estimates of the high-energy physicists are fully justified in this
more sophisticated (and realistic) model. We note, however, that the
interest in the polarization of constituents of \Lipol6 is not limited
to this one investigation in high-energy physics, but indeed opens up
possibilities of a variety of experimental and theoretical
investigations of the spin structure of \Lipol6. Some of these will be
discussed in the closing section of this paper.

But is the three-body model of \Li6 truly realistic?  The accumulating
evidence from the confrontation of calculations with experiment
answers this question with an emphatic yes. Many properties of the
trio of $A=6$ nuclei at low excitation energies ($\leq$ 15 MeV) can be
understood within the context of exact three-body theory and good
phenomenological representations of the low energy (below 23 MeV)
behavior of the basic interactions \cite{Lehad,Lehshell}. The quality
of the predictions is better than that of effective two-cluster models
or the standard shell model which reduces in its effect to two-body
dynamics. As two recent examples of this claim, we cite predictions
of the proton-knockout reaction \Li6$(e,e'p)$ and of the
quadrupole moment $Q$. The former reaction is well described below
$\alpha$ breakup by a three-body model and cannot be described by a
(small space) shell model calculation \cite{Lanen}. According to the
Erlangen group \cite{hofmann}, the latter small negative value of $Q$
cannot be reproduced in a cluster model by a pure ($\alpha\rm D$) wave
function. Instead a modern resonating-group model needs a three-body
($\alpha n p$) wave function with $s$ and $d$ partial waves between
the clusters and within the ``deuteron'' to match experiment. This
result corroborates the earlier conclusions of the Surrey group and
others \cite{surrey}. These predictive aspects of the three-body
model are even more interesting once one realizes that, in contrast to
the effective two-body models, once the parameters of the model are
determined at the $\alpha$N and NN level, no further parametrization
is allowed and all the results obtained thereafter are direct
predictions of the model.

In this paper we present solutions of the Faddeev equations for a
number of three-body models of \Li6 which differ by various
combinations of $\alpha$N and NN potentials. For each model, we
calculate the longitudinal charge form factors $F_{C0}$ and $F_{C2}$
and associated lowest moments (charge radius and quadrupole moment) to
assess the static properties of the wave function. From the wave
function we calculate the polarization of the neutron and the proton
outside the $\alpha$. We find for all cases a polarization of the
neutron in excess of 90\% of the polarization of the \Li6 nucleus,
implying an effective neutron polarization in excess of 30\%.  This
degree of polarization is not as high as the predicted 87\%
polarization of the neutron in a fully polarized $^3$He target
\cite{Friarpol}. However, for some nuclear physics purposes, a {\em
dense} (\Li6D is a solid), {\em isoscalar} ($T=0$) polarized target
with an effective neutron polarization of 45\% (the polarization of
the neutron in $\roarrow{\rm D}$ is over 95\%) may be an attractive
alternative to obtaining polarized neutrons from a {\em gaseous},
$T=\tfrac1/2$, $^3\roarrow{\rm He}$ target.

We note already the first use of a polarized \Li6 target for nuclear
research to obtain the angular distribution of the vector analyzing
power $i T_{11}$ for $\pi^+$-\Lipol6 elastic and inelastic scattering
\cite{tacikpion}. We will discuss the interpretation of this
experiment in a later section, but note for the moment that
polarization predictions of models of \Li6 should be tested by
knockout reactions \cite{Lanen,lampf1}, scattering by polarized proton
beams \cite{lampf2}, and other experimental probes. That is, a
program for \Li6 similar to the program which tests the $^3$He wave
function by knockout reactions \cite{rahav}, scattering by polarized
proton beams \cite{CE25}, etc.\ would, in our opinion, be helpful to
test our predictions. This is, firstly, because the prediction is
couched as the answer to the question ``If we pick a nucleon from the
fully polarized nucleus without disturbing its spin, what is the
degree of polarization of that spin?''  This question is a theorist's
question and may or may not be answerable in a given experiment. For
example, any spin-dependent final-state interactions or spin-dependent
meson-exchange currents can alter the ``plucked'' nucleon's spin.
Secondly, the answer is based on calculated wave-function
probabilities (or, in the general case with isospin breaking,
off-diagonal matrix elements) which are unambiguously obtained from
the interaction operators. These quantities are not experimental
observables, however, because of the problems of defining uniquely the
relativistic corrections to the interaction operators
\cite{Amado,Friar}. Hence, experimental tests of our predictions are
required. We hope that our predictions could be used to motivate such
experiments.

The paper is organized as follows: Sec.\ \ref{Sec:Formalism} briefly
sketches our solution to the Faddeev equations. We derive expressions
for constituent polarization based on wave function probabilities in
Sec.\ \ref{Sec:Polarization}. After a brief description of the
potentials which differentiate the models in Sec.\
\ref{Sec:Potentials}, we discuss the different sources of uncertainty
within the three-body model (by looking at binding energies, charge
radii, quadrupole moments) in Sec.\ \ref{Sec:Uncertainty}. A
comparison of observables with experiment and other theoretical
calculations to establish the quality of the three-body model itself
is given in Sec.\ \ref{Sec:Quality}. Our predictions for polarization
of the valence nucleons and other aspects of wave function
probabilities are discussed in Sec.\ \ref{Sec:NuclPol}. We give a
summary and outlook in Sec.\ \ref{Sec:Summary}.


\section{Formalism} \label{Sec:Formalism}

We will be modeling \Li6 as a three-body cluster model, assuming the
$\alpha$ particle to be structureless. The three-body problem will be
solved by applying the spline method to the configuration-space
Faddeev equations. The form of the Faddeev equations is well known and
will simply be sketched here. We will, however, give some definitions
which are essential of the understanding to the rest of this paper.

The Schr\"odinger equation for three particles interacting via
two-body forces
\begin{equation}
  (H_0 + V_1 + V_2 + V_3 - E)\Psi = 0
 \,,
\end{equation}
where $V_i$ is the interaction between particles $j$ and $k$, and
$H_0$ is the kinetic energy operator, is equivalent to the Faddeev
equations
\begin{eqnarray}
 \nonumber
 (E-H_0-V_1)\psi_1 & = & V_1 (\psi_2+\psi_3)
  \\ \nonumber
 (E-H_0-V_2)\psi_2 & = & V_2 (\psi_3+\psi_1)
  \\
 (E-H_0-V_3)\psi_3 & = & V_3 (\psi_1+\psi_2)
 \,,
\end{eqnarray}
from
which the total wave function can be constructed by addition of the
Faddeev amplitudes $\psi_i$:
\begin{equation}
  \label{TotalPsi}
  \Psi = \psi_1 + \psi_2 + \psi_3
 \,.
\end{equation}
Another, sometimes useful form of the Faddeev equations is
\begin{equation}
  \label{FaddEqAlt}
  (E-H_0)\psi_i = V_i\Psi \,,
\end{equation}
which follows immediately from Eq.~(\ref{TotalPsi}).

By expanding the Faddeev amplitudes on a basis of angular-momentum
eigenfunctions (also known as channels), the Faddeev equations can be
rewritten as an infinite set of coupled two-dimensional partial
(integro-)differential equations. This set can be approximated by a
finite subset, which can be solved numerically. For this purpose we
use the spline method \cite{nico}.

The $\alpha$ particle will be labeled as particle one, the nucleons as
particles two and three. We will ignore the isospin-symmetry breaking
in most cases, and assume a pure $T=0$ state for the \Li6 nucleus. We
will perform one calculation which includes explicit isospin breaking,
to check if this simplification is justified. The breaking is
generated by including the Coulomb force, and considering the nucleons
to have different masses. (We will consider particle two to be the
neutron, and particle three to be the proton.)

For further discussion of the Faddeev equations in the notation above,
and the numerical method used to solve them we refer to Ref.\
\cite{nico}. A more up-to-date description of our methods will be
published at a later time. Now we will discuss the angular momentum
basis functions used in our calculations in some detail, since it is
both enlightening and instructive.

The three Faddeev equations are numbered using the labels of the
noninteracting, or {\em spectator} particles, {\em i.e.}, Faddeev
equation $i$ has particle $i$ as the spectator particle. The orbital
momentum of the two interacting particles relative to each other in
Faddeev equation $i$ is denoted by $l_{x_i}$, and the orbital momentum
of the pair with respect to the spectator particle by $l_{y_i}$. (Note
that for every Faddeev equation there is a natural Jacobi coordinate
system. The corresponding Faddeev amplitude will be expressed in this
particular coordinate system.)

For the pure isosinglet case two of the three Faddeev equations are
dependent, so that we only have to solve two independent equations.
Consider equation one, {\em i.e.}, the equation in which the $\alpha$
particle is the spectator. Antisymmetry of the wave function demands
that $l_{x_1}+s_{23}+t_{23}$ be odd (note that $t_{23}=T$), parity
demands that $l_{x_1}+l_{y_1}$ be even. For the second Faddeev
equation there is only the parity requirement. We will use $jj'$
coupling, {\em i.e.}, the coupling scheme
$((l_{x_i}s_{jk})j_{x_i}(l_{y_i}s_i)j_{y_i})JM_J$. The Faddeev
equations are solved including all channels which have $j_{x_i} \leq
j_{\rm max}$, where $j_{\rm max}$ is varied to check the convergence.
For the largest calculation we have used $j_{\rm max}=8.5$,
corresponding to $75$ channels. For the case including isotriplet
admixture we have to solve all three Faddeev equations. Also, the
antisymmetry requirement is no longer present since we consider the
two nucleons to be distinguishable, so that the number of channels
doubles when compared to the corresponding isosinglet case.

Note that our calculations are complete in the sense that only states
with very high two-body angular momenta are left out. This is a
notable difference from most calculations in the literature which
employ far fewer channels. The dependence of our results upon the
number of channels and comparison with other calculations will be
deferred until Sec.\ \ref{Sec:Uncertainty}.

To investigate the quality of the wave function, we study the
longitudinal form factors and the associated moments as an example of
the static properties of the wave function. The charge radius and
quadrupole moment are given by
\begin{eqnarray}
 \braket<r^2> ^{1/2} & = & (\tfrac2/3\braket<r_1^2 >
 + \tfrac1/3\braket<r_2^2> )^{1/2} \,,
\end{eqnarray}
\begin{eqnarray}
 \braket<Q > & = & \sqrt{\tfrac16\pi/5}
  (2\braket<r_1^2Y_{20}(\hat{r}_1)>   +
   2\braket<r_2^2Y_{20}(\hat{r}_2)>)
 \,.
\end{eqnarray}
They are related to the low-energy limit of the Coulomb form factors
\begin{eqnarray}
 \label{Eq:FormFactor}
 &   & F_{Cl}(q) \nonumber \\
 & = & \frac{\sqrt{4\pi}}{Z} \sum_i Z_i
\braket<J||Y_l(\hat{r}_i)j_l(qr_i)||J>  \nonumber \\
 & = & \frac{\sqrt{4\pi}}{Z} \sum_i
       \frac{Z_i}{\braket<J\,J\,l\,0|J\,J> }
       \braket<J\,J|Y_{l0}(\hat{r}_i)j_l(qr_i)|J\,J>
 \,,
\end{eqnarray}
where $Z=3$ and $J=1$ for \Li6. The low-energy limit of $F_{C0}$ is
\begin{eqnarray}
 &   &
  \frac{\sqrt{4\pi}}{Z} \sum_i
  \frac{Z_i}{\braket<1\,1\,0\,0|1\,1> }
  \braket<1\,1|Y_{00}(\hat{r}_i)j_0(qr_i)|1\,1> \nonumber \\
 & \longrightarrow &
  \frac{1}{Z} \sum_i Z_i \braket<1\,1|1-\tfrac1/6(qr)^2|1\,1> \nonumber \\
 & = &
  1-\tfrac1/6\braket<r^2> q^2
 \,.
\end{eqnarray}
The low-energy limit of $F_{C2}$ is
\begin{eqnarray}
  \label{Eq:Lowqlimit}
 &   &
  \frac{\sqrt{4\pi}}{Z} \sum_i
  \frac{Z_i}{\braket<1\,1\,2\,0|1\,1> }
  \braket<1\,1|Y_{20}(\hat{r}_i)j_2(qr_i)|1\,1> \nonumber \\
 & \longrightarrow &
  \frac{\sqrt{4\pi}}{Z} \sum_i
  \frac{Z_i\sqrt{10}}{(2\cdot2+1)!!}
  \braket<1\,1|Y_{20}(\hat{r}_i)(qr)^2|1\,1> \nonumber \\
 & = &
 \frac{1}{3Z\sqrt{2}}Qq^2
 \,.
\end{eqnarray}

The inclusion of the Pauli principle will be discussed in
Sec.~\ref{Sec:Potentials}. In the next section we will concentrate
on the polarization of constituents.


\section{Polarization of constituents} \label{Sec:Polarization}

The question we want to ask is the following: if we pick a nucleon
from the fully polarized nucleus ($J_Z=J$) without disturbing its
spin, what are the odds for finding its spin ``up''. A similar
description of polarization was first used in Ref.~\cite{Friarpol} for
the case of $\rm{}^3He$.

We will follow a similar line of reasoning here; first for the case of
the deuteron to establish notation and then for \Li6.
The quantity
\begin{equation}
 P_n^+  =  \braket<J\,J|\hat{P}_n^+|J\,J>
\end{equation}
where
\begin{eqnarray}
 \hat{P}_n^+ & = & \frac{1-\tau_3(1)}{2}\frac{1+\sigma_3(1)}{2}
                 + \frac{1-\tau_3(2)}{2}\frac{1+\sigma_3(2)}{2} \nonumber \\
             & = & P_n(1)P^+(1)+P_n(2)P^+(2)
\end{eqnarray}
and $\ket|J\,J> = \ket|\Psi(J_Z=J)>$, is the probability for the
neutron to have its spin aligned ``up''. Due to the fact that the
deuteron is in a pure $J=1$, $T=0$ state, as is \Li6 in models without
isospin-symmetry breaking, the two nucleons have to be in the
isosinglet state. This means that there is only one isospin function
in the wave function, which can be factored out. This simplifies the
calculations significantly, as we find
\begin{eqnarray}
 P_n^+ & = &
 \hskip-1mm\hphantom{+} \braket<T=0|P_n(1)|T=0>\braket<J\,J|P^+(1)|J\,J>
\nonumber \\
       &  &
 \hskip-1mm+ \braket<T=0|P_n(2)|T=0>\braket<J\,J|P^+(2)|J\,J>
\nonumber \\[1mm]
      & = &
  \tfrac1/2\braket<J\,J|P^+(1)|J\,J> + \tfrac1/2\braket<J\,J|P^+(2)|J\,J>
 \,.   \label{eq:jj}
\end{eqnarray}
The remaining matrix elements can be calculated by separating the spin
and orbital parts using a simple case analysis. First we consider the
deuteron which has the orbital momentum states $\{S,D\}$ coupled to
the spin triplet state. Since spin projection operators leave orbital
angular momenta invariant and there is exactly one spin state per
orbital momentum state, there is no coupling between spin states, so
that we find for the deuteron the diagonal terms:
\begin{eqnarray}
  \braket<11|P^+(i)|11 > & = & P_S \braket<(01)11|P^+(i)|(01)11> \nonumber \\
                         &  & + P_D \braket<(21)11|P^+(i)|(21)11>
 \,.   \label{eq:angmom}
\end{eqnarray}
where $P_S$ and $P_D$ are the probabilities ($P_S + P_D = 1$) of the
orbital angular momentum states $\ket|(LS)JJ>$ of the deuteron. The
remaining matrix elements of (\ref{eq:angmom}) are of angular momentum
functions only. They can be evaluated using:
\begin{eqnarray}
      &   & \braket<(LS)JJ|P^+(i)|(LS)JJ> \nonumber \\
      & = & \sum_{MM'} \braket<L\,M\,S\,M'|L\,S\,J\,J>^2
            \braket< S\,M'|P^+(i)|S\,M' >
 \,. \nonumber
\end{eqnarray}
Evaluating these matrix elements yields the simple and well-known
result:
\begin{equation}
  \label{eq:pdeuteron}
 P_n^+  =   P_S + \tfrac1/4 P_D \,,
\end{equation}

Now we turn to the polarization of the two ``valence'' nucleons of
\Li6 since in our model the $\alpha$ particle is an elementary
particle, and its constituents are therefore outside our grasp. If at
first we neglect the isotriplet component of the wave function, the
isospin function can be factored out and the possible orbital momentum
states of the three-body system are $\{S,P,P',D\}$, where a prime is
used in an unconventional way: to denote a state with odd $l_{x_1}$.
The $S$, $P$, and $D$ states are coupled to the spin triplet state,
the $P'$ is coupled to the spin singlet state. (Just as for the
isospin sum, the absence of spin structure in the $\alpha$ particle
simplifies matters enormously.)  Again there is no coupling between
spin states and we find:
\setbox0=\hbox{$P_{P'}$}
\begin{eqnarray}
  \braket<11|P^+(i)|11 > & = & \hphantom{+}
          \hbox to\wd0{$P_S$\hfil} \braket<(01)11|P^+(i)|(01)11> \nonumber \\
 &    & + \hbox to\wd0{$P_P$\hfil} \braket<(11)11|P^+(i)|(11)11> \nonumber \\
 &    & + P_{P'} \braket<(10)11|P^+(i)|(10)11> \nonumber \\
 &    & + \hbox to\wd0{$P_D$\hfil} \braket<(21)11|P^+(i)|(21)11>
 \,.
\end{eqnarray}

The final result is very simple, and in contrast to the analog formula
for $\rm{}^3He$, it is exact:
\begin{equation}
    \label{eq:plithium}
 P_n^+  =  P_S + \tfrac3/4 P_P + \tfrac1/2 P_{P'} + \tfrac1/4 P_D \,.
\end{equation}
If, for some reason, only the total $P$ state probability $P_{P_{\rm
TOT}}$ is known, we still have the following bound:
\begin{equation}
	\label{eq:boundlithium}
 P_S + \tfrac1/2 P_{P_{\rm TOT}} + \tfrac1/4 P_D
  \leq  P_n^+
  \leq  P_S + \tfrac3/4 P_{P_{\rm TOT}} + \tfrac1/4 P_D \,.
\end{equation}
We remark that from the corresponding analysis for the proton it
follows that $P_n^+ = P_p^+$ in this model of $T=0$ \Li6 (and in the
deuteron).

The Coulomb interaction does not conserve isospin so that a \Li6 model
with the Coulomb force is not a pure $T=0$ state but has a small
admixture of $T=1$. Because there is more than one isospin state there
can be more than one spin state per angular momentum state.  This
means that the diagonal probabilities cannot be used in all cases. The
analysis outlined above is unchanged for the $S$ and $D$ states,
however, because the spin triplet state cannot couple to the spin
singlet (since there is no spin singlet component in these states).
Hence the $S$- and $D$-state contributions to $P^+_n$ can still be
written in terms of probabilities. For the $P$ and $P'$ states the
$T=1$ admixture allows more than one spin state per orbital momentum
state and the best one can do is derive the following bound in terms
of the total $P$-state probability:
\begin{equation}
 P_S + \tfrac1/4 P_{P_{\rm TOT}} + \tfrac1/4 P_D
  \leq  P_n^+
  \leq  P_S + P_{P_{\rm TOT}} + \tfrac1/4 P_D \,.
\end{equation}
We note that this bound is not as tight as the (unnecessary) bound
derived above for the isospin-symmetric model. Finally, although the
{\em bounds} are the same for the polarization of the neutron and the
proton,
\begin{equation}
 P_n^+  \neq  P_p^+ \,.
\end{equation}
in this most realistic three-body model of \Li6. In practice, the
difference is small because $P_{P_{\rm TOT}} \approx 3\%$.

It is important to emphasize that the formulae derived above for the
polarization of the constituents of \Li6 are intended to develop an
intuition for our numerical results. The values of $P^+_n$ presented
in the tables for the various Hamiltonians did not come from these
formulae but were calculated exactly in the codes.


\section{$\alpha$N and NN potentials} \label{Sec:Potentials}

We will neglect the microscopic structure of the $\alpha$ particle.
Its structure is partially represented by the $\alpha$N interaction
and by its charge form factor. The $\alpha$N potentials we will use
are of a phenomenological nature: they more or less fit the $\alpha$N
low-energy phase shifts. However, most potentials support a
deeply-bound $\alpha$N state, which contradicts experimental
observation. The justification for this seeming discrepancy is
thought to be the Pauli principle which excludes this bound state.

The Pauli principle can be taken into account in several
essentially different ways. First, one can use a potential that is
repulsive in the $s$ state, so that no forbidden bound state is
supported. Second, one can suppress the forbidden bound state, by
replacing the Hamiltonian $H$ by:
\begin{equation}
 H'  =  PHP
 \,,
\end{equation}
where $P$ is an operator that projects out the forbidden states:
\begin{equation}
 P  =  (1-\ketbra|\psi_2><\psi_2|)(1-\ketbra|\psi_3><\psi_3|)
 \,,
\end{equation}
where $\ket|\psi_i>$ is the forbidden two-body bound state of
particles $j$ and $k$. Another possibility is to suppress not the
Pauli-forbidden state itself, but some harmonic-oscillator state close
to it.

The relative merits of the two different methods have been studied
phenomenologically for some time now. The experimental properties of
the three-body system appear not to distinguish between these choices
\cite{Lehfs,Lehmom}. Supersymmetric quantum mechanics provides a new
tool for studying the application of the Pauli principle to scattering
of a projectile from a composite system. That is, it is always
possible to generate from a given local deep (shallow) potential a
corresponding shallow (deep) potential which produces the same phase
shifts \cite{Baye}. The scalar Hamiltonians H0 and H2 associated with
these ``phase equivalent'' potentials can be represented as components
of a supersymmetric Hamiltonian which act in subspaces of an overall
supersymmetric space \cite{Sukumar}. These subspaces are analogous to
the bosonic and fermionic sectors in supersymmetric formulations of
field theory \cite{Witten}. Given H0(H2) we can derive H2(H0) by means
of a supersymmetric transformation. This has been done for an
$\alpha\alpha$ interaction \cite{Baye} and for the $\alpha$N
interaction \cite{ssqm}. The phase equivalent shallow potential has a
repulsive $r^{-2}$ singularity for small $r$ in agreement with the
notion that Pauli effects for composite systems are repulsive in the
range of wave function overlap. Although the supersymmetric
transformation provides a unique local potential that gives the
scattering and has no Pauli-forbidden bound state, to our knowledge,
there are no few-body calculations which probe the off-shell
properties of the supersymmetric partners. This subject and the
different methods for handling the problem of unphysical $s$-wave
bound states generated by $\alpha$N potentials will be discussed in
detail in a subsequent publication.

The method we have used in our calculations is a variation of the
second approach. We replace the $\alpha$N potentials $V_i$ by:
\begin{equation} \label{Eq:PauliSuppression}
 V_i'  =  V_i+\Gamma\ketbra|\psi_i><\psi_i| \,,
\end{equation}
and let the (positive) constant $\Gamma$ go to infinity. We will show
that this limit can be taken analytically, and that in the limit this
method is exactly equivalent to solving the three-body equation in the
restricted space.

The resolvent for the two-body Schr\"odinger operator
\begin{equation}
 H'  =  H_0+V+\Gamma\ketbra|\psi_f><\psi_f| \,,
\end{equation}
can be written as
\begin{eqnarray}
 \nonumber
 G'(E) = G(E)-\frac{\ketbra|\psi_f><\psi_f|}{(E-E_f)(1-\frac{E-E_f}{\Gamma})}
 \,,
\end{eqnarray}
where $G(E)$ is the resolvent when $\Gamma=0$, $\ket|\psi_f>$ a
Pauli-forbidden state, and $E_f$ its bound-state energy. Using this
expression in Faddeev equation $i$ by embedding it in three-body
space, we find:
\widetext
\begin{eqnarray}
 \ket|\psi_i> & = & \left(G_i
       -\int\!{\rm d}^3\vec{q}_i
        \frac{\ketbra|\vec{q}_i\psi_{if}><\vec{q}_i\psi_{if}|}
             {(E-E_{if}-q_i^2)(1-\frac{E-E_{if}-q_i^2}{\Gamma})}\right)
        V'_i (\ket|\psi_j>+\ket|\psi_k>)
 \,,
\end{eqnarray}
where $\vec{q}$ is the spectator momentum. When $\Gamma$ is large, we can
approximate this by:
\begin{eqnarray}
 \ket|\psi_i> & = & \left(G_i
       -\int\!{\rm d}^3\vec{q}_i
        \frac{\ketbra|\vec{q}_i\psi_{if}><\vec{q}_i\psi_{if}|}{E-E_{if}-q_i^2}
        [1+\Gamma^{-1}(E-E_{if}-q_i^2)+O(\Gamma^{-2})]\right) \nonumber \\
 & & \times
 \left(V_i+\Gamma\int\!{\rm d}^3\vec{q}_i'
      \ketbra|\vec{q}_i'\psi_{if}><\vec{q}_i'\psi_{if}|\right)
 (\ket|\psi_j>+\ket|\psi_k>)
 \nonumber \\
 & = & G_i (V_i+\Gamma P_f) (\ket|\psi_j>+\ket|\psi_k>)
      -\int\!{\rm d}^3\vec{q}_i
       \frac{\ketbra|\vec{q}_i\psi_{if}><\vec{q}_i\psi_{if}|}
            {E-E_{if}-q_i^2}
       (V_i+\Gamma) (\ket|\psi_j>+\ket|\psi_k>) \nonumber \\
 & & - P_f(\Gamma^{-1}V_i+1) (\ket|\psi_j>+\ket|\psi_k>)+O(\Gamma^{-1})
 \,,
\end{eqnarray}
\narrowtext\noindent
where we have written $P_{if}$ to denote
\begin{equation}
  \int\!{\rm d}^3\vec{q}_i \ketbra|\vec{q}_i\psi_{if}><\vec{q}_i\psi_{if}|
 \,. \\
\end{equation}
(Note that $P_{if}^2 = P_{if}$, and that $[P_{if},G_i] = [P_{if},H_i]
= 0$.) Using the fact that
\begin{equation}
  G_iP_f = \int\!{\rm d}^3\vec{q}_i
           \frac{\ketbra|\vec{q}_i\psi_{if}><\vec{q}_i\psi_{if}|}
                {E-E_{if}-q_i^2} \,, \\
\end{equation}
we finally arrive at:
\begin{eqnarray}
  \label{NewFadEq}
 \ket|\psi_i> & = & G_i (1-P_{if}) V_i (\ket|\psi_j>+\ket|\psi_k>)
                  - P_{if} (\ket|\psi_j>+\ket|\psi_k>) \nonumber \\
              &   & + O(\Gamma^{-1})
 \,.
\end{eqnarray}
To prove that this equation gives the correct solution in the limit
$\Gamma\to\infty$, we must show that (in this limit)
\begin{equation}
  (1-P_{if})(E-H)\ket|\Psi> = P_{if}\ket|\Psi> = 0 \,. \\
\end{equation}
Multiplying Eq.~(\ref{NewFadEq}) by $P_{if}$, we find:
\begin{equation}
 P_{if}(\ket|\psi_i> + \ket|\psi_j> + \ket|\psi_k>) = P_{if}\ket|\Psi> = 0 \,,
\end{equation}
This proves the first condition, provided that Eq.~(\ref{TotalPsi})
defines the correct Schr\"odinger wave function. Now we multiply
Eq.~(\ref{NewFadEq}) with $(E-H_i)$:
\begin{eqnarray}
 &   & (E-H_i)\ket|\psi_i> \nonumber \\
 & = & (1-P_{if}) V_i (\ket|\psi_j>+\ket|\psi_k>)
      -(E-H_i)P_{if}(\ket|\psi_j>+\ket|\psi_k>) \nonumber \\
 & = & V_i (\ket|\psi_j>+\ket|\psi_k>)
      -P_{if}(E-H_0)(\ket|\psi_j>+\ket|\psi_k>)
 \,.
\end{eqnarray}
Now use Eq.~(\ref{FaddEqAlt}) to eliminate $\ket|\psi_j>$ and
$\ket|\psi_k>$ from the right-hand side:
\begin{equation}
 (E-H_i)\ket|\psi_i> = V_i(\ket|\psi_j>+\ket|\psi_k>)-P_{if}(V_j+V_k)\ket|\Psi>
 \,,
\end{equation}
which, when added to the two remaining Faddeev equations gives
\begin{equation}
 (E-H_0)\ket|\Psi> = -P_{if}(V_j+V_k)\ket|\Psi> \,.
\end{equation}
We now multiply with $(1-P_{if})$ to complete our proof:
\begin{eqnarray}
 (1-P_{if})(E-H_0)\ket|\Psi>           & = & \nonumber \\
 (1-P_{if})(E-H_0)(1-P_{if})\ket|\Psi> & = & 0 \,,
\end{eqnarray}
showing that the second condition we needed is satisfied, and that
$\ket|\Psi>$ is indeed a solution of the Schr\"odinger equation in the
restricted space, as we claimed earlier.

We will now discuss the potentials actually used by us in this paper.
The nucleon-nucleon potentials used are: the Reid soft core potential
(RSC) \cite{RSC}, the super soft core potential (SSC) \cite{SSC}, and
the ($s$-wave) Malfliet-Tjon I-III potential (MT) \cite{MT}. The RSC
and SSC potentials are realistic nucleon-nucleon interactions, and
have a tensor force. The MT potential is a simple model without a
tensor force, but it is very popular in the literature. The
alpha-nucleon potentials used are: the simple Sack-Biedenharn-Breit
potential (SBB) \cite{SBB} of which we use three versions, a potential
with the Wood-Saxon form of Satchler {\em et al.}\ (SAT)
\cite{Satchler}, and one of the potentials obtained on a grid from an
inversion of proton-alpha scattering data by Cooper and Mackintosh
(CM) \cite{CM}.

There is considerable confusion over the parameters of simple model
potentials. The exact parameters of the potentials we use are given in
Table \ref{Tab:PotParm}. The algebraic forms are:
\[
 V_{\rm MT}  =  V_1 \frac{e^{-\mu_1 r}}{r} + V_2 \frac{e^{-\mu_2 r}}{r} \,,
\]
for the Malfliet-Tjon NN potential \cite{MT},
\[
 V_{\rm SBB} =  V_1 e^{-(r/\mu_1)^2}
          + (\vec{l}\cdot\vec{s}) V_2 e^{-(r/\mu_2)^2} \,,
\]
for the Sack-Biedenharn-Breit $\alpha$N potential \cite{SBB}, and
\[
 V_{\rm SAT} =  V_c
          + (\vec{l}\cdot\vec{s})\frac{1}{r}\frac{{\rm d}}{{\rm d}r}V_{ls} \,,
\]
where
\[
 V_{c,ls} =  \frac{V_1}{1+e^{(r-\mu_1)/\mu_2}} \,,
\]
for the Satchler $\alpha$N potential. The optical model potential of
Satchler {\em et al.}\ \cite{Satchler} has traditionally been modified for
three-body calculations by neglecting the energy dependence of $\mu_1$
\cite{kuk1,BG} and refitting the parameters. We make the same
approximation and use the parameters of Refs. \cite{kuk1,BG}.

Both potentials SBB and SAT have the same radial form for odd- and
even-parity potentials, in contrast to effective local potentials
which are deemed equivalent to the nonlocal interaction obtained from
resonating group model (RGM) calculations (see, for example, Fig.\ 26
of \cite{Ali}). The source of the parity dependence in RGM
calculations is exchange processes. To examine this odd-even feature
of $\alpha$N potentials, we also use the CM potential which is
strongly parity dependent; having short-range even-parity potentials
and long-range odd-parity potentials. The energy dependence of the CM
potential, obtained by an inversion procedure, is rather less that
that of SAT which arises from a direct optical model fit of the
data. We make the usual energy independent approximation (which
should be better for the CM potential) by choosing the 12 MeV version
as shown in Fig.\ 4 of Ref.\ \cite{CM}

The RSC and SSC potentials are as defined in \cite{RSC} and
\cite{SSC}, respectively. We used the exact physical masses for all
constituents. For systems without isospin breaking, we used the
average mass of the proton and the neutron as the nucleon mass.
We used two different Coulomb potentials. The first, taken from
\cite{PP}, is
\[
 V = \left\{ {
      {\displaystyle \frac{2e^2}{2R}(3-(r/R)^2) \qquad r  <   R}
      \atop
      {\displaystyle \frac{2e^2}{r}             \qquad r \geq R}
     } \right.\,,
\]
where $R = 1.25\cdot4^{1/3}\rm\,fm$. The second, taken from \cite{kuk1} is
\[
 V = \frac{2e^2}{r}{\rm erf}(r/a) \,,
\]
where $a = \sqrt{\frac23}\cdot1.64\rm\,fm$.


\section{Sources of uncertainty} \label{Sec:Uncertainty}

In this section we will discuss the results obtained with our method
for various models and the various sources of uncertainty. These can
be separated into numerical uncertainties, which will be discussed
first, and the uncertainties due to uncertainties of the input: the
interactions between the constituents. The validity of the three-body
model will be not be discussed here, but we will comment on it in the
next section.

\subsection{Numerical approximations}

There are three types of numerical approximations in our calculations,
one of which can (and will be) avoided, and two which are unavoidable.
The approximations are: ({\em i\/}) the forbidden state is not
projected out but suppressed using a separable term as in
Eq.~(\ref{Eq:PauliSuppression}), ({\em ii\/}) the spline
approximation, the accuracy of which is determined by the number of
intervals in which the domain of the partial differential equation is
subdivided, and ({\em iii\/}) the partial-wave series is approximated
by a finite number of partial waves. (Note that for potentials defined
on a finite number of partial waves, the results are exact if all
these partial waves are included in the calculation.) The errors
associated with the approximations ({\em i\/}) and ({\em ii\/}) reduce
systematically as the suppression parameter $\Gamma$ and the number of
intervals in the grid are increased. This is illustrated by Tables
\ref{Tab:GammaConv} and \ref{Tab:SplineConv}. The error associated
with approximation ({\em iii\/}) is not as easily expressed as a
function of the relevant parameter ($j_{\rm max}$), so that only very
crude extrapolations can be made, and it is essential to use a very
high number of channels. This is shown in Table \ref{Tab:ChanConv}.
It appears that $j_{\rm max}=6.5$ is sufficient to calculate all the
observables with a precision that is better than or comparable to that
of experiment.

For most models, it is possible to make $\Gamma$ extremely large
$O(10^8)$, so that we actually reach ``infinity'' for all practical
purposes. Another possibility is to take the limit analytically in the
manner we have shown earlier. This was done for almost all
calculations. Table \ref{Tab:GammaConv} shows that all the observables
converge nicely to the value for $\Gamma=\infty$, and do so with
leading $\Gamma^{-1}$ behavior, as expected. (Note that $\Gamma$ is
expressed in units of $\hbar^2/M$, where $M$ is approximately one
nucleon mass. The exact value of $\hbar^2/M$ is $41.47\rm\,MeV\,fm^2$.)

As is well known \cite{nico}, the error of a spline approximant is
$O(h^4)$ (where $h$ is the length of an interval in the grid),
provided $h$ is sufficiently small. This fact can be exploited, by
taking suitable linear combinations of results obtained for different
grid sizes, which effectively means extrapolating to an infinitely
fine grid. The observed deviations from fourth-order convergence are
used to estimate the error.

We would like at this moment to stress a very interesting point, which
is often neglected and has led to confusion. It is illustrated by
Table \ref{Tab:ChanConv}: the Faddeev eigenvalue $E_F$ and the
expectation value of the Hamiltonian $\braket<H>$ differ
substantially. This difference decreases when the number of channels
increases. The explanation for this phenomenon is the following:
Solving the Faddeev equations for a certain set of partial waves is
exactly equivalent to solving the Faddeev equations for all partial
waves if the two-body potentials $V_i$ are restricted to operate in
these partial waves only. Therefore, if we evaluate $\braket<H^*>$
where $H^*$ is the Hamiltonian in which the potentials are restricted
to operate in the set of partial waves for which we solved the Faddeev
equations, we will find it to be equal to $E_F$.

However, the expectation value of the full Hamiltonian $H$ will be
different from $\braket<H^*>$. This can be understood as follows: the
total (Schr\"odinger) wave function is the sum of the three Faddeev
amplitudes. These three amplitudes can be written using a finite
number of channels, provided the expansion is done in the natural
(Jacobi-) coordinate system of each amplitude. After adding the
amplitudes this is no longer possible, since a state that has good
quantum numbers $l_x$ and $l_y$ in one coordinate system does in
general not have corresponding good quantum numbers in any other
system. The total wave function therefore contains an infinite number
of angular momentum components. The full Hamiltonian operates in all
channels, so that there will be a contribution to $\braket<H>$ from
these {\em induced} channels.

Usually, the potentials are attractive. This means that $\braket<
H>$ will be below $\braket<H^*>$. In other words: it will
be closer to the exact (ground-state) energy. This is why a wave
function constructed from Faddeev amplitudes with a certain number of
channels is usually closer to the full wave function than a wave
function obtained from direct solution of the Schr\"odinger equation
for the same number of channels.

Extrapolation to an infinite number of channels is very difficult, and
we have used it very conservatively. We are confident that the errors
listed (the number in parentheses is the uncertainty in the last
quoted digit) are not too small. (A confirmation of this can be found
in Table \ref{Tab:ChanConv}, by comparing the extrapolated values of
$E_F$ and $\braket<H>$: the difference between the two is well within
the error bars.)  The rest of the results presented in this paper are
the result of extrapolation to infinite grid size and an infinite
number of channels, {\em i.e.}, they are absolute predictions of \Li6
observables in the three-body model.

\subsection{Models}

We have performed full calculations for eight different interaction
models. The models used are: ($a$) RSC+SBB, ($b$) RSC+SBB with
isospin breaking, ($c$) RSC+$\rm SBB_2$, ($d$) RSC+$\rm SBB_3$, ($e$)
RSC+SAT, ($f$) RSC+CM, ($g$) SSC+SBB, and ($h$) MT+SBB. Note that
only model $b$ has isospin breaking. We will also display partial
results for some of these models where the strong interaction is
augmented by the Perey-Perey Coulomb force (but ignoring its isospin
breaking). These results are labeled ($a^*$) RSC+SBB, ($e^*$)
RSC+SAT, ($f^*$) RSC+CM, ($g^*$) SSC+SBB, and ($h^*$) MT+SBB.

Models $a$, $a^*$, and $b$ are used as reference models, since they
describe the Lithium observables very well (provided the Coulomb force
is included, either directly as in $a^*$ and $b$, or by perturbation
theory). Note that we assume the SBB potential to interact in all
partial waves, since we feel that a realistic interaction cannot have
interaction in the lowest partial waves only. Model $c$ contains the
SBB potential in a somewhat modified form ($\rm SBB_2$), as suggested
by Bang {\em et al.}\ \cite{Bohrinst}. The modifications are twofold:
({\em i\/}) the $d$-wave interaction is attenuated to better fit the
phase shifts, ({\em ii\/}) the potential only works in $s$, $p$, and
$d$ waves. Model $d$ contains a third version of the SBB potential
\cite{Danilin1}, which replaces the attractive Gaussian $s$-wave
potential of $\rm SBB_2$ by a repulsive one and therefore does not
support a forbidden state. In this $\rm SBB_3$ the $p$-wave and
$\vec{l}\cdot\vec{s}$ parameters are those of the original SBB, the
$d$-wave strength is attenuated, and, of course, the $s$-wave
potential is repulsive. Models $e$ and $f$ retain the RSC potential
as the NN interaction but enlarge the range of alpha-nucleon
interactions studied. Finally models $g$ and $h$ return to the SBB
alpha-nucleon potential but vary instead the nucleon-nucleon
interaction. Parameters of the potentials we use are given in Table
\ref{Tab:PotParm} and in Refs. \cite{RSC} and \cite{SSC} for the RSC
and SSC nucleon-nucleon potentials. Note that the values in Table
\ref{Tab:PotParm} may differ slightly from the original references.

\subsection{Model dependence}

We will now discuss the effect of variations of different parameters
in our model. These are ({\em i\/}) the manner in which the Pauli
principle is taken into account in the alpha-nucleon interaction,
({\em ii\/}) the presence or absence of isospin-symmetry breaking,
({\em iii\/}) different forms of the $\alpha\rm N$ interactions, and
({\em iv\/}) different forms of the NN interactions.

Looking at Table \ref{Tab:ModelOverview} and comparing models $c$ and
$d$ we see that there is little difference between the two models. In
the case of \Li6 it hardly matters whether potentials supporting
Pauli-forbidden states in the $s$ wave ({\em e.g.}\ $c$) or repulsive
potentials ({\em e.g.}\ $d$) are used. The only difference seems to be
that the attractive potential binds \Li6 marginally less strong. This
appears to be a confirmation of the results of Lehman
\cite{Lehfs,Lehmom} obtained with separable $\alpha\rm N$ potentials.

Comparing models $a$, $a^*$, and $b$ we see that although there is
(probably because \Li6 is so lightly bound) substantial
isospin-symmetry breaking (the effect on the binding energy is about
$0.15\rm\,MeV$), it appears that only the binding energy and the
charge radius are substantially affected. The other observables are
hardly changed at all. It therefore seems justified to ignore the
$T=1$ components, provided that one bears in mind that a noticeable
energy effect is to be expected.

The influence of the form of the $\alpha\rm N$ interaction is rather
large, as Table \ref{Tab:ModelOverview} shows. This is to be expected
since there is not nearly as much consensus on this interaction as
there is on the NN interaction. This is due to the composite structure
of the alpha particle, the amount of experimental data available, and
finally the number and quality of $\alpha\rm N$ model interactions
available. The use of high-precision inversion techniques is a rather
recent development. The structure of the alpha particle, and the
amount of data available make the inversion rather difficult, since
there is a large amount of nonlocality (which can be expressed as
energy dependence or dependence on the angular momentum) and
uncertainty in the interaction.

It is somewhat disappointing to see that the rather crude potentials
SBB and SAT reproduce the \Li6 binding energy much better than the
more sophisticated CM potential, especially since this potential
yields an attractively low value for the quadrupole moment $Q$ (we
will discuss the sign of the quadrupole moment later on). It is known
that the ground-state energy of \Li6 in a three-body model is not
sensitive to phase shifts at higher energy of the $\alpha$N
interaction and that low-energy phase shift properties of the
potential are the most important for this purpose \cite{Lehold}. We
checked the $s$ wave phase shifts of the three potentials and found
that the SBB and SAT reproduced the threshold behavior \cite{arndtlow}
much better than did the CM potential. We may speculate that the CM
potential placed a greater emphasis on an overall fit to the phase
shifts than to getting the low energy behavior very well. It is
possible that a slight refit of the CM potential which emphasizes
threshold behavior could yield a much better \Li6 model.

We found the two $\alpha p$ Coulomb models to be hardly
distinguishable: the Coulomb energy for the Perey-Perey model was
systematically about $8\rm\,keV$ larger than that for the
error-function model. This is a negligible difference when compared to
the uncertainties encountered earlier.

Often, the Coulomb potential is treated as a perturbation. Looking at
Table 5, we see that the difference of the ground-state energy plus
Coulomb energy of a system and the ground-state energy of this system
in which the Coulomb potential is taken into account exactly are about
$5\rm\,keV$ (except for the very lightly bound system $f$, there the
difference is about $27\rm\,keV$), indicating that first-order
perturbation theory for the Coulomb potential is a very good
approximation of the $T=0$ approximation of \Li6 (even though the
Coulomb energy is relatively large, when compared to other nuclear
systems), and therefore a fairly good approximation of the full
system, yielding errors of a few percent (caused mainly by symmetry
breaking, which is a higher-order effect).

Comparing models $a$, $g$, and $h$, which share the SBB $\alpha$N
interaction and vary the NN interaction, we see that the influence of
the NN interaction on most observables is very small. However, model
$h$ has a substantially lower value for $Q$. This is due to the
absence of a tensor force in this model (which implies much lower $P$-
and $D$-state probabilities). Note also, that as a consequence of
this, the predicted polarization of the valence nucleons is
significantly larger than for the other models.

\subsection{Summary}

Summarizing, we think that within the three-body model: ({\em i\/})
the numerical uncertainties in this work are negligible, ({\em ii\/})
ignoring the Coulomb potential in the Hamiltonian does not invalidate
the general predictions that can be made, and ({\em iii\/}) by far the
greatest source of uncertainty is the $\alpha$N potential. (However,
general properties are described rather well, provided the potentials
fit the low-energy scattering data.)


\section{Quality of the three-body model} \label{Sec:Quality}

Having established the main source and the magnitude of uncertainty
within our model, we continue with a discussion on the quality of the
three-body model itself by comparing with experiment and with other
calculations found in the literature.

\subsection{Comparison with experiment}

In Fig.~\ref{Fig:ScatterPlot} we have plotted the charge radius as a
function of the binding energy, and find that the points scatter
around a line, which goes through the experimental datum. We find this
very encouraging, since it appears that three-body models can
accurately describe the general features of the \Li6 nucleus. As
Fig.~\ref{Fig:ScatterPlot} shows, it is only possible to draw such a
conclusion about a fit to the experimental datum, if a sufficiently
large number of channels is used. If the number of channels is too
small, the charge radius will be automatically too large. This effect
can be substantial even when the energy appears to have converged,
since an error of magnitude $\varepsilon$ in the wave function will
result in an error of $O(\varepsilon^2)$ in the binding energy and an
error of $O(\varepsilon)$ in the other observables. This is more or
less in agreement with the suggestion by Bang and Gignoux \cite{BG}
who claim their binding energy has converged although they use a small
number of channels. We remind the reader that the results presented
in Fig.~\ref{Fig:ScatterPlot} and the latter tables are the result of
extrapolation to infinite grid size and an infinite number of
channels, {\em i.e.}, they are absolute predictions of \Li6
observables in the three-body model.

Comparing the results in Table \ref{Tab:ModelOverview} for models {\em
a} and {\em b}, we find that the Coulomb force does not play a very
important role in \Li6, apart from lowering the binding energy by
approximately $0.9\rm\,MeV$, and increasing the charge radius
accordingly. In Fig.~\ref{Fig:FormFactor} we show the longitudinal
form factors up to a four-momentum transfer of $5\rm\,fm^{-1}$ for
models {\em a} and {\em b}. We used the impulse approximation of
Eqs.~(\ref{Eq:FormFactor}--\ref{Eq:Lowqlimit}) modified by folding in
the nucleon \cite{Hohler} and alpha-particle \cite{ffhe4}
electromagnetic form factors. Our calculation predicts the
experimental data \cite{ffli6} rather well up to about
$q=2.5\rm\,fm^{-1}$. The good agreement with experiment reflects the
accuracy of our charge radius for these models and the fact that the
dominant monopole term swamps the effect of the quadrupole term in
this region. However, since the zero in the C2 contribution occurs at
somewhat higher $q$ than that of the C0, we expect the agreement to be
less good in the region of the diffraction minimum at about
$3.0\rm\,fm^{-1}$.

Our quadrupole form factor is close to that obtained by Eskandrian
{\em et al.}\ \cite{lehff}, and rather different from that obtained by
Kukulin {\em et al.}\ \cite{kuk3}. We find that the form factors for
model {\em a} and {\em b} are very close to each other, again
confirming that the Coulomb potential is not very important to the
charge form factor. Our result does not support the suggestion by
Kukulin {\em et al.}\ \cite{kuk3} that the disregard of the Coulomb
force by Bang and Gignoux and by Lehman and coworkers is the cause of
their overestimate of the theoretical C0 form factor in the region up
to the first minimum. A recent preprint from Kukulin {\em et al.}
\cite{kuk92} demonstrates that much of the disagreement of \cite{kuk3}
with Bang and Gignoux, with Lehman and coworkers, and with us, on C0
and C2 (and many other predictions of their wavefunction) was caused
by a severe truncation in partial waves of their variational wave
function. Their truncation difficulties will be discussed further on
in this section, but here we note that the new \cite{kuk92} C2 form
factor from that group is consistent with
Fig.~\ref{Fig:FormFactor}.

Figure \ref{Fig:FormFactor} shows the absolute prediction of the
longitudinal form factor of \Li6 for our interaction models $a$ and
$b$ which are described by a neutron, a proton, an $\alpha$ particle,
and their interaction operators. We have not attempted to calculate
meson-exchange contributions (MEC) to this ``impulse approximation''
in order to be consistent with our other predictions of observables
and with our predictions of constituent polarizations. These latter
predictions also drop corrections of relativistic order. This is
because there are interaction-dependent ambiguities which arise from
separating a completely relativistic result (which we do not have)
into a nonrelativistic part plus corrections. For us the most
analogous nucleus to \Li6 is the isoscalar-vector deuteron where these
ambiguities in calculating the MEC to the $q\rightarrow 0$ charge
radius and quadrupole moment are not large \cite{klarsfeld} and the
corrections themselves are small. At higher momentum transfer, the
MEC become larger and the interaction-dependent ambiguities also
become of greater concern. One particular prescription which ties
much but not all of the MEC to the interaction operator, indicates
that the pion seagull contribution is largest and of opposite sign to
the dominant impulse approximation for $F_L$ of the deuteron
\cite{riska}. That is, the diffraction minimum is shifted to slightly
lower $q$ and the second maximum is enhanced by MEC. If one could
confidently apply these (model-dependent) deuteron results to the case
of \Li6, one would conclude that our result of an impulse approximation
form factor slightly above the data for higher $q$ is further
confirmation of the essential correctness of our models $a$ and $b$.
For this paper, we simply state that our calculated $F_L$ is a direct
prediction of the interaction model and that it agrees with the data
at low momentum transfer.

To conclude the discussion of Table \ref{Tab:ModelOverview}, we
briefly look at the $\alpha$-D clustering probability, and the
Coulomb energy. Although these are not observables, they do provide a
crude consistency test: one expects scaling between the charge radius
and the Coulomb energy and (since the binding energy as well as the
clustering probability mainly depend on the strength of the $\alpha$D
interaction) scaling between the binding energy and the clustering
probability. Table \ref{Tab:ModelOverview} confirms these
expectations. The results are also in reasonable agreement with those
found in the literature.

We will now turn to the magnetic moment, which is the $q\to0$ limit of
the transversal form factor. We will use the observation of Lehman
{\em et al.}\ \cite{lehmm} and others that the magnetic moment for the
ground state of \Li6 can be written in terms of probabilities:
\begin{equation}
   \label{eq:magmom}
 \mu = \mu_p + \mu_n
     + (\tfrac1/2 - \mu_p - \mu_n) [\tfrac1/2 P_P + P_{P'} + \tfrac3/2 P_D] \,.
\end{equation}
This simple formula has been critized by Danilin {\em et al.}\
\cite{Danilin1} as arising from a shell model which does not make
allowance for the motion of the $\alpha$ particle, but they also show
that the $\alpha$ orbital motion gives a contribution of 0.1\% to the
magnetic moment. Meson exchange current contributions to the magnetic
moment are isoscalar and therefore of order $v^2/c^2$ \cite{Friar83}.
Such relativistic corrections should be consistently neglected in
calculations such as ours. In any event, we calculate the magnetic
moments of our models with (\ref{eq:magmom}) and display the results
in Table \ref{Tab:ProbsOverview} and Fig.~\ref{Fig:MuPol}. The
straight-line relationship between $\mu$ and $P_n^+$ is inherent in
the formulas and introduces no new results. However, it is a test of
self-consistency. (Note that model $b$ is slightly off the dashed
line. This is caused by the fact that the neutron polarization in
model $b$ cannot be expressed as wave-function probabilities, due to
isospin breaking. If we were to plot the average polarization of the
neutron and the proton for this model, it would again be exactly on
the line.) It is gratifying to see that models $a$ and $b$ which have
binding energies, longitudinal form factors, and charge radii in
excellent agreement with experiment also predict a magnetic moment
within $1\%$ of the experimental value of $0.82205$ nuclear magnetons.

\subsubsection*{The quadrupole moment}

So far, the results are very encouraging. Let us now turn to a
well-known problem in three-body models of \Li6: the sign of the
quadrupole moment. As far as we know no dynamical three-body model has
ever produced a negative value for $Q$, as is demanded by experiment.
Could the experimental value be somehow wrong, or does this
discrepancy point out a serious deficiency of the dynamical three-body
model?

The experimental value of the quadrupole moment of \Li6 has changed
since many of the phenomenological \cite{surrey}, resonating group
\cite{hofmann}, and dynamical three-body model results were published.
The new value is still very small and negative (indicating a slightly
oblate nucleus with a primarily equatorial distribution of charge) but
is now known to be about 30\% more negative than the early value of
$-0.064\,e\,{\rm fm}^2$ still found in compilations and review
articles. Quadrupole moments of nuclei are deduced from the
experimental quadrupole coupling obtained from molecular resonance
spectroscopy. From such experiments the quadrupole moment can only be
extracted if a reliable theoretical value of the electric field
gradient at the nucleus in the molecule is available. The experimental
quadrupole coupling from the molecule \Li6$\rm^1H$ can only be
obtained with error bars of over $10\%$. Therefore one uses the ratio
$Q(\Li6)/Q(\Li7)$ of $+0.0205(20)$ from experiments on LiF to deduce
the value of $Q(\Li6)$ from the by now well determined value of
$Q(\Li7)$ \cite{Qexp}. A recent analysis of both molecular
spectroscopy and Coulomb-scattering experiments reviews the last
twenty years of effort and arrives at the consistent value of $Q(\Li7)
= -4.00\pm0.06\,e\,{\rm fm}^2$ \cite{Fick}. Applying the ratio, one
arrives at a value of $Q(\Li6) = -0.082\,e\,{\rm fm}^2$ to compare
with theoretical results.

This value is rather small when compared with $Q({\rm D})=+0.2860 \pm
0.0015\,e\,{\rm fm}^2$ \cite{BC} and explanations have been attempted
in terms of $\alpha\rm D$ cluster models. For an orientation to our
results, we repeat the arguments of Ref.~\cite{hofmann} which begin by
mentioning that a two-cluster ($\alpha\rm D$) wave function with zero
relative angular momentum would have no electric quadrupole moment and
an additional $d$ wave on the relative motion would give
$Q(\Li6)\simeq -1\,e\,{\rm fm}^2$. A three-cluster wave function,
however, consisting of an alpha particle, a proton, and a neutron with
$s$ and $d$ waves between them would have a positive $Q$ because of
the positive $Q$ of the deuteron. An interplay between the $s$ and $d$
waves ($l_y$) between the clusters and within the deuteron ($l_x$) was
found to account for the experimental $Q(\Li6)$ in a resonating group
calculation. (This is the same argument as those of
Refs.~\cite{surrey}, except that a deuteron with $S$- and $D$-waves is
counted as a single cluster in those references).

Turning now to our calculations (Table \ref{Tab:ModelOverview}), we
find that none of the models has a negative value for the quadrupole
moment, implying that all our models render a slightly prolate \Li6.
Since there are no adjustable parameters we must accept this. The
present calculations seem to contradict the suggestion (made by Bang
and Gignoux \cite{BG}, as well as Kukulin {\em et al.}\ \cite{kuk3})
that the correct sign for $Q$ can be obtained, just by taking a
sufficient number of channels. The feeling that small wave function
components play a decisive role in the value of $Q$ is justified (cf.\
Table \ref{Tab:ChanConv}), since $Q \propto
\braket<3z^2>-\braket<r^2>$ is a ``difference'' operator: its value is
the result of a subtraction of two, almost equal, numbers. For a
(nearly) spherical wave function we have that $\braket<z^2>$ is nearly
1/3 times $\braket<r^2>$, so that $Q/\braket<r^2> \approx
\braket<3\cdot(1/3)\cdot(r^2+\varepsilon)> / \braket<r^2>-1 =
\varepsilon/\braket<r^2> \ll 1$. This makes $Q$ extremely sensitive
to small details in the wave function. As it turns out, however, the
small wave components tend to make $Q$ more positive and further away
from experiment. Danilin {\em et al.}\ \cite{Danilin1} suggest that
the experimental value of $Q$ for \Li6 may be evidence for an
intrinsic polarization of the $\alpha$ particle in the field of a
valence proton and would correspond to a quadrupole charge deformation
parameter $\beta_0$ of the order of 10\%. This refinement is outside
of the scope of our model.

However, there is still room for improvement inside the three-body
model. It must be noted that most $\alpha$N model potentials fit
low-energy data moderately well, and high-energy data not at all. One
may argue that this is not very significant since it appears that the
low-energy behavior is dominant in the ground state of \Li6
\cite{Lehold}. However, since $Q$ is very sensitive to small details,
we cannot ignore this shortcoming.

We have also found that the value of $Q$ is sensitive to the
angular-momentum structure of the $\alpha$N interaction and the
strength of the tensor force in the NN interaction. For example, the
Malfliet-Tjon I-III NN potential with no tensor force predicts
$Q(\Li6_{TH}) = +0.26\, e\,{\rm fm}^2$, even though the ``deuteron'' in
the wave function has no $D$ wave. Evidently, the interplay between
the $s$ and $d$ waves ($l_y$) between the clusters and within the
deuteron ($l_x$) demanded by potentials which fit the phase shifts is
different from that which produces a negative value of $Q$. Another
interesting case is model {\em f}. It is the only model that has a
Majorana component in the $\alpha$N interaction and it has a
quadrupole moment which is significantly smaller than the other models
(forgetting about models {\em c}, {\em d}, and {\em h}, which must be
considered unrealistic). Unfortunately, model {\em f\/} fails to bind
\Li6 sufficiently. This seems to be correlated to a rather poor fit to
low-energy scattering data of this potential. In our opinion there is
an urgent need for a truly energy-independent $\alpha$N potential,
which fits the scattering data over a large energy range. (Such a
potential would have to be $l$ dependent.)

The quadrupole moment may be a feature of \Li6 which is sensitive to
an often neglected feature of dynamical three-body models and the
defect may be cured by antisymmetrizing the wave function. Note that
the three-body model is not fully antisymmetric, since exchange of
nucleons inside the $\alpha$ particle with the valence nucleons has
only approximately been treated. For example, an analysis of the
charge form factor by Unkelbach and Hofmann \cite{UH} finds that both
C0 and C2 contributions are very sensitive to the effects of
antisymmetrization for their three cluster wave functions. These
authors use the resonating group method \cite{RGM} and their wave
functions are therefore fully antisymmetric with respect to an
exchange of any two of the six nucleons. Upon full antisymmetrization
C2 takes on the correct sign at low $q$ and displays the unusual
feature that the second maximum is larger than the first; both effects
presumably due to cancellations of direct and exchange contributions
generated by antisymmetrization.

It would appear, however, from another resonating group calculation
which included more angular momentum channels \cite{Attila},
that the negative value of $Q$ obtained in \cite{hofmann} might be a
truncation artifact; this more recent RGM calculation finds $Q(\Li6) =
+0.25\, e\,{\rm fm}^2$, but $Q$ becomes negative when they truncate to
(almost) the model of Ref.\ \cite{hofmann}. A recent preprint by
Kukulin {\em et al.} \cite{kukanti} with variational wave functions in
a greatly expanded space displays a C0 unchanged by antisymmetrization
and the minimum of C2 moved to slightly smaller $q$.

On the other hand, Hofmann finds that reducing the RGM model space
does not change the sign of $Q$, but a separate calculation with a
fully antisymmetrized wave function does give the correct sign
\cite{Hofmanncomm}. We suggest that further work is needed to resolve
this discrepancy and mention that the formalism needed to fully
antisymmetrize the (converged) Faddeev wave functions of the dynamical
three-body model is being developed \cite{Adam}.

\subsubsection*{Summary}

To summarize the results of this subsection, we believe that the
three-body model gives a quantitative description of \Li6 properties,
except for the quadrupole moment. We therefore believe that it is
justified to use this model as a starting point for the determination
of polarization of the constituents.

\subsection{Comparison with other results}

In this subsection we ignore experiment and compare with other
calculations within the same model. A theorist's concern is then ({\it
i\/}) do the results accurately reflect the input potentials (issues
of convergence etc.), and ({\it ii\/}) how do the results depend on
the input potentials. These questions were addressed in the previous
section with regard to our calculations. In this subsection, we will
concentrate on the results found in the literature in relation to our
results. The data presented in Fig.\ \ref{Fig:ScatterPlot2}, Table
\ref{Tab:ModelOverview}, and Table \ref{Tab:Compar} will now be
discussed with these questions in mind. Table \ref{Tab:Compar}
presents a selection of the results found in the literature with
different calculational methods (coordinate-space variational
\cite{kuk1,kuk92,kuk2}, coordinate-space direct solution of the
Faddeev equations \cite{BG}, hyperspherical expansion \cite{Danilin1},
and Faddeev integral equations in momentum space \cite{Lehold,lehmm})
applied to a variety of input potentials.

Because the \Li6 radii of the variational calculations have already
been noticed to be somewhat at odds with expectations from the binding
energies \cite{lehff}, and because the wave functions of
Ref.~\cite{msuwf} have been used to analyze pion-scattering from \Li6
\cite{tacikpion}, we attempted to reproduce the results
Ref.~\cite{kuk1}. As can be seen in Table \ref{Tab:Compar} (and in
Fig.\ \ref{Fig:ScatterPlot2}), a Faddeev calculation for model $b$ with
RSC NN potential and SBB $\alpha$N potential with (approximately) the
same partial waves as those of \cite{kuk1,kuk2}, and finite
suppression parameter $\Gamma$ gives a binding energy about 2\%
smaller than our converged result and the full radius is 6\% larger
than the converged result. (This model is denoted by $a^*_{\rm Kuk}$
because the truncation of partial waves only allows $T=0$ and
therefore it has no isospin breaking.)  It is somewhat surprising,
then that the result of Ref.~\cite{kuk1}, listed as Kukulin-84, has a
{\em full} charge radius which 10\% {\em smaller} than our converged
result with this model. The explanation of this discrepancy may lie
(as already noted in Ref.\ \cite{lehff}) in the fact that the true
wave function has, of course, the correct exponential falloff at large
distances, and the Gaussian wave functions they utilize do not.
However, the close agreement between $a^*_{\rm Kuk}$ and Kukulin-84
for both the energy and the (bare) radius, we think it is more likely
that the radii listed in Table 1 of Ref.~\cite{kuk1} are bare,
although Kukulin claims to list full radii
\cite{KukPrivComm}.

Continuing the comparison of Kukulin-84 and our model $b$ truncated to
$a^*_{\rm Kuk}$, we find that both models give a value of
$Q\approx+0.2$. However, we find a much larger $P$-state probability
for this model the authors of \cite{kuk1} do. This probability, while
not observable, does play a role in constituent polarizations,
magnetic moments, and pion scattering as will be discussed shortly.
Based on these comparisons, and the earlier discussion of longitudinal
form factors and the charge radii, we believe that the wave functions
from this early variational approach do not accurately reflect the
input potentials. The same can be said for the results denoted by
Kukulin-86 in Table \ref{Tab:Compar}, where both the $P$- and
$D$-state probabilities appear to be too small. The variational
results from a greatly expanded number of angular momentum channels
\cite{kuk92}, labeled Kukulin-92 in Table \ref{Tab:Compar}, are
another story. They are in reasonable agreement with our results for
the same model $b$.

The other examples of calculations in Table \ref{Tab:Compar} are those
of Bang and Gignoux \cite{BG}, Danilin {\em et al.}\ \cite{Danilin1},
and Lehman {\em et al.}\ \cite{lehff,lehmm}. None of the inputs
exactly match our models $a$--$h$, so the comparisons can only be
qualitative. As mentioned in the discussion of our results, the
convergence of the binding energy with the small number of channels
used by Bang and Gignoux is not so bad and their wave function results
are in qualitative agreement with those of models $e$, and $g$, the
closest to their choice of potentials. The hyperspherical-expansion
results \cite{Danilin1} with yet another smooth NN potential
(Gogny-Pires-de Toureil \cite{GPT}) and the Sack potential are
admitted to not have fully converged. But various methods are used to
extrapolate to their asymptotic results in Table \ref{Tab:Compar}
which are in qualitative agreement with those of Table
\ref{Tab:ModelOverview}. The integral equations of
\cite{lehff,lehmm} have to account only for mesh sizes and not a
truncation of partial waves. This appears to be well under control as
discussed in their papers and as shown by the nice straight line fit
in the radius-energy plane (cf.\ Fig.\ \ref{Fig:ScatterPlot2}). As
mentioned earlier, this is not as much the case for the results from
\cite{kuk1}. Especially the point $(B,r)=(3.393,2.35)$, corresponding
to Kukulin-84 and our model $a^*_{\rm Kuk}$ strikes us as odd. Most
recently, their ``final'' result for this model (obtained with the
RSC in all NN channels and a greatly expanded model space) is
$(B,r)=(3.33,2.44)$. Note also that one of the results of \cite{kuk1}
lies above the $\alpha\rm D$ breakup threshold, and therefore should
not have a well-defined radius.

After checking on the reliability of the results a theorist turns to
the question of how does the output depend on the input. The most
systematic attempt to answer this second question is found in the
program of Lehman and collaborators who fit five different separable
potentials to the {\em same} low energy NN and $\alpha$N data. Each
set of potentials addresses a specific question such as the strength
of the tensor force in the NN interaction (labeled by 0\% or 4\% $D$
state in the deuteron as shown in Table \ref{Tab:Compar}) or whether
the $s$-wave $\alpha$N interaction does support a state forbidden by
the Pauli potential that must be projected out (``proj.'') or does not
(``rep.''). In our opinion, the drawback of these potentials is that
they do not contain a short-range repulsion, they do not include the
negative parity terms of the NN interaction, and they are separable
and the nonlocality is hard to relate to an underlying meson-exchange
picture. Finally, the Coulomb force cannot be included in the model.
Thus we feel that there is room for further studies such as ours.

One aspect of the dependence of the three-body model results upon the
input which has entered into the folklore has to do with the wave
function probability $P_{P'}$. It is stated in its purest form in a
recent publication \cite{Attila}: ``It is plausible and has been
corroborated by test calculations that the weight of this contribution
depends crucially and almost solely, on the singlet odd $^1P_1$ term
of the nucleon-nucleon force [$\ldots$]'' This assertion perhaps stems
from the statement in \cite{kuk1} that the value of $P_{P'} = \beta^2
\ll 1\%$ was due to the repulsive NN interaction in the $P$-wave: ``If
the NN interaction in the $P$-wave is excluded, the value of $\beta$
will rise.''  However, a better numerical treatment of the RSC
potential soon raised $P_{P'}$ from $0.27\%$ in \cite{kuk1} to the
$1.08\%$ quoted in \cite{kuk2} and Table \ref{Tab:Compar} and
eventually a large expansion of the angular momentum channels lead to
the 2.12\% of the ``final'' solution Kukulin-92 \cite{kuk92} displayed
in Table \ref{Tab:Compar}. The results in Kukulin-92 are in
reasonable agreement with our results for the same model but $P$-state
probabilities are still somewhat smaller than the other calculations.
Now we return to the relation between $P_{P'}$ and the NN interaction
in the $^1P_1$ channel. It is well known that the $^1P_1$ phase
shifts of the RSC potential are about half the size of experiment,
because Reid fitted his potential to phase shift solutions from data
which has since been discarded \cite{Brady,SAID}. Perhaps we should
be concerned because we used a bad NN potential for models $a$--$f$.
Perhaps, but not for this reason. Notice that in Table
\ref{Tab:ProbsOverview} the probability $P_{P'}$ in \Li6 varies by a
factor of six for the same RSC NN potential and different $\alpha$N
potentials, whereas it hardly changes when RSC (model $a$) is replaced
by SSC (model $g$) which has better $^1P_1$ phase shifts. That is,
our results do not support this bit of \Li6 folklore.


\section{Nucleon polarization in \mbox{$\bf^6\roarrow{\bf L\lowercase{i}}$}}
\label{Sec:NuclPol}

The predictions of the polarization of the neutron, expressed as a
percentage of the fully polarized \Lipol6, have already been displayed
in Table \ref{Tab:ModelOverview}. In this table, the calculated
polarizations are greater than 90\% for the valence nucleons in the
three-body model of \Li6, close to an estimate of 97\% obtained from
the two-body model of \Li6 with the aid of (\ref{eq:pdeuteron}).
However, the overlap of the three-body wave function on the ``$\alpha
+ \rm D$'' structure gives a probability of 60--70\% for most models
so the naive guess from the two-body model of \Li6 must be scaled down
to about 58--68\%. It is remarkable that the ``uncorrelated'' valence
nucleons of the part of the wave function which is not ``$\alpha + \rm
D$'' in the exact calculation boost this back up to $P^+_n\geq 90\%$.
In Table \ref{Tab:ProbsOverview} we give the probabilities of the
wave-function components and the calculated value of $P^+_n$ for our
models. Note that the proton polarizability is marginally larger than
the neutron polarizability for model $b$, the only model which has a
$T=1$ component. For the other models there is just one
polarizability, since the proton and neutron are considered
identical. It is clear from (\ref{eq:plithium}) that $P^+_n \ge P_S$
and that the exact value of the neutron polarization depends on the
smaller probabilities of the $P$ and $D$ states. Since these are well
defined within a model but are not observables \cite{Amado,Friar},
polarization predictions of models of \Li6 should be tested by
knockout reactions \cite{Lanen,lampf1}, scattering by polarized proton
beams \cite{lampf2}, and other experimental probes. That is, a program
for \Li6 similar to the program which tests the $^3$He wave function
by knockout reactions \cite{rahav}, scattering by polarized proton
beams \cite{CE25}, etc., would, in our opinion, be helpful to test our
predictions.

In the absence of such a program, we can use the fact that the
magnetic moment for the ground state of \Li6 can also be written in
terms of these probabilities, as discussed in the previous section, to
help calibrate our polarization estimates. Observing as we did earlier
that models $a$ and $b$ which have binding energies, longitudinal form
factors, and charge radii in excellent agreement with experiment also
predict a magnetic moment within 1\% of the experimental value of
0.82205 nuclear magnetons, we feel there is strong support to the
polarization prediction of 93\% from these models.

The magnetic moment is the zero momentum transfer limit of the
transverse form factor. This form factor has been invoked in attempts
to understand the angular distribution of the vector analyzing power
$iT_{11}$ for $\pi^+$-\Lipol6 elastic and inelastic scattering. The
argument goes as follows. ``The transversal form factor $F_T$ is
expected to be very close to [the pion-nuclear spin form factor]
$F^S_{el}(q)$ because the convection current contribution almost
vanishes in $e$-\Li6 scattering, if $L=0$ is really the dominating
configuration in the \Li6 ground-state wave function''
\cite{tacikpion}. This statement can be quantified at the $q\to0$ limit
by the explicit formula for the orbital contribution $\mu_L$ to the
magnetic moment \cite{lehmm}
\begin{equation}
 \mu_L = \tfrac1/2 [ \tfrac1/2 P_P + P_{P'} + \tfrac3/2 P_D]  \,.
\end{equation}
Typical values from Table \ref{Tab:ProbsOverview} suggest that the
orbital motion contributes $\sim 0.05$ to the total moment of $\sim
0.83$. In the existing treatments of pion-nucleus scattering, the
details of the pion-nucleon interaction are subsumed into this spin
form factor $F^S_{el}(q)$ whose shape is determined by the underlying
nuclear wave function. The analysis of \cite{tacikpion} finds that a
shell-model wave function describes the data marginally better than
one of the earliest variational three-body wave functions of Kukulin
{\em et al.}\ \cite{msuwf}. Our probabilities of the $P$- and
$D$-state components from Table \ref{Tab:ProbsOverview} (typically
$P_{P'}\approx 2$--$3\%$, $P_P \approx 0.2$--$0.6\%$, $ P_D \approx
6.5$--$6.7\%$) are quite different from those of the wave function
\cite{msuwf} ($P_{P'}\leq 0.5\%$, $P_P = 0.0$, $ P_D =0.0 $) used in
\cite{tacikpion}. The spin form factor from any of the wave functions
of our models is quite likely to be different from the three-body
model already used. At present we can only speculate on the changes
which might occur if one of our wave functions (or indeed one from
Ref.\ \cite{kuk92}) were used to analyse the pion scattering data. A
new calculation of the angular distribution of the vector analyzing
power $i T_{11}$ for $\pi^+$-\Lipol6 elastic and inelastic scattering
would seem to be in order before one could conclude finally that the
three-body model of \Li6 cannot explain this data as well.


\section{Summary and outlook} \label{Sec:Summary}

We have solved the configuration-space Faddeev equations of the
dynamical three-body model which characterizes \Li6 as a bound system
of an alpha particle, a neutron, and a proton interacting with local
potentials which parametrize the free space forces between the three
particles. We have made an exhaustive study of the convergence of our
solutions both in the spline approximation and in the number of
partial waves (channels) kept in the angular-momentum expansion of the
Faddeev amplitudes. Our observables (and wave fuction expection
values) are the result of extrapolation to infinite grid size and an
infinite number of channels, {\em i.e.}, they are absolute
predictions of the model. For this reason, we were able to compare
with earlier solutions, which did not include so many channels, to
draw conclusions about the quality of the concomitant wave functions.

Most of our models used the Reid soft core NN potential with a variety
of $\alpha$N potentials which more or less fit the low-energy
$\alpha$N phase shifts. We could estimate the effect of the Coulomb
potential perturbatively or could include it in the full Hamiltonian
and found that perturbation theory works very well. Futhermore, the
Coulomb potential has a tiny effect on all expectation values except
the binding energy and the charge radius. All the models assumed a
pure $T=0$ state for \Li6 except one (labeled $b$) which checked this
simplification by including explicit isospin breaking generated by the
Coulomb potential and the neutron-proton mass difference. Isospin
breaking increased the binding energy by about 4\%. Most $\alpha$N
potentials support $s$-wave states in the five-nucleon system which
are forbidden by the Pauli principle. These Pauli-forbidden states
are projected out exactly using a new equation for the Faddeev
amplitude which we have shown to be the numerical limit of the
traditional pseudopotential approach used to deal with forbidden-state
potentials \cite{Lehfs,kuk1}.

Comparison of observables of our converged calculations with the
experimental properties of the ground state of \Li6 was quite
satisfying. For thirteen models a plot of charge radius versus
binding energy (Fig.\ \ref{Fig:ScatterPlot}) displays points which
scatter around a straight line which passes through the experimental
datum. Our most sophisticated models $a^*$ and $b$ (RSC NN potential,
a gaussian potential by Sack {\em et al.}\ which fits $s$- and
$p$-wave phase shifts of the $\alpha$N interaction, and the Coulomb
interaction without and with isospin breaking) are within 10\% of the
experimental binding energy and within the error bars of the charge
radius. The experimental longitudinal form factor is described well
up to momentum transfer $q=2.5\rm\,fm^{-1}$. The magnetic moment is
described well by most of the models and extremely well by the models
which are near the radius-binding energy datum. The very small
quadrupole moment is not described well by our calculations, nor by
any calculation in the dynamical three-body model of \Li6 . The
explanation of this discrepancy may well lie outside this model.

Our motivation for undertaking these calculations was to examine,
within a well-defined and controlled model, the extent that the
nucleons in \Lipol6 are themselves spin aligned and could therefore be
considered polarized targets for measurements of the spin structure
function of the neutron and proton. It is, of course, the polarized
neutron which is the desired target because it can only be utilized
when bound in a nucleus. \Lipol6 has been suggested as a target for
the study of direct photon production with polarized proton beams and
polarized nucleon targets which could determine the spin-dependent
gluon distribution in the nucleon \cite{fermilab}. This information
on the uncharged constituents of the nucleon is needed to complement
the knowledge of the spin-dependent quark and antiquark distributions
obtained from deep inelastic scattering of longitudinally polarized
leptons from longitudinally polarized nuclei. The first results from
lepton scattering on the spin-dependent structure function $g_1(x)$ of
the deuteron have been published \cite{SMC}, and those from $^3$He
have been announced \cite{SLAC}. To infer from this data the first
moment of the spin-dependent neutron structure function (for a
comparison with sum rules), one needs to know the polarization of the
nucleons within the polarized nucleus. Our equation
(\ref{eq:pdeuteron}) for the deuteron and similar formulae for $^3$He
\cite{Friarpol} provide this information theoretically. In this paper
we have developed the formulae and done the calculation for the
polarization of the valence nucleons of the three-body model of \Li6.
We find a polarization of the neutrons in excess of 30\% of the
\Lipol6.  This theoretical result implies that \Li6D should provide a
very good target of polarized neutrons (45\%) for the hadronic
experiments to determine the polarized gluon distribution of the
nucleon.

Our result is couched as the answer to the question ``If we pick a
nucleon from the fully polarized nucleus without disturbing its spin,
what is the degree of polarization of that spin?''  This question is a
theorist's question and may or may not be answerable in a given
experiment. Indeed, we have not seen a discussion of this point in
the few published results of polarized lepton scattering and wish to
reiterate the necessity of testing such predictions before extracting
{\em neutron} spin-dependent structure fuctions from measurements on
nuclei. For example, the HERMES experiment will measure the
spin-dependent structure functions of hydrogen, deuterium, and $^3$He
in one series of measurements \cite{HERMES}. In addition, there is a
vigorous experimental program to support the deep inelastic scattering
measurements which tests the $^3$He wave function by knockout
reactions \cite{rahav}, scattering by polarized proton beams
\cite{CE25}, etc. In our opinion, no less a commitment should be
made to the interpretation of the proposed direct photon production
experiments at Fermilab \cite{fermilab}. That is, our polarization
predictions of models of \Li6 should be tested by knockout reactions
\cite{Lanen,lampf1}, scattering by polarized proton beams
\cite{lampf2}, and other experimental probes.


\acknowledgments

This project began when SAC and NWS came together at the Institute for
Nuclear Theory of the University of Washington. We wish to thank the
Institute Director Ernest Henley and Program Director Jim Friar for
creating a stimulating and productive atmosphere at the Institute. It
was also supported in part by National Science Foundation grant
\hbox{PHY-9017058}. NWS wishes to thank the physics department of New
Mexico State University for hospitality and, in particular, the
geophysics group for generously providing computer support during a
short visit. SAC acknowledges stimulating and helpful discussions
with Ana Eir\'o, Antonio Fonseca, Teresa Pe\~na, Shirley Cooper, Ray
Mackintosh, Ron Johnson, George Burleson, Gary Kyle, and the members
of his nuclear theory class at NMSU. We have benefited from e-mail
discussions with Hartmut Hartmann and V.I. Kukulin and from the
receipt of preprints from the latter. RMA would like to thank FOM
(Fundamenteel Onderzoek der Materie) for financial support and Ben
Bakker for useful discussions.

\raggedright


\begin{figure}
\caption{%
The charge radius of \Li6 versus the binding energy $B$ for breakup
into $\alpha$+$n$+$p$. The converged results of the models with
different local potentials are labeled as in the text. Open circles
are models without the Coulomb interaction and the solid symbols
include the Coulomb interaction. The solid circles labeled ``$a^*$''
include Coulomb but assume a pure $T=0$ state. Our one complete
calculation which included isospin breaking of the Coulomb force is
the filled triangle. Also shown is the convergence of results of a
single model (model $a$) with the number of channels. The straight
line through our points is to guide the eye; it is the result of a
least squares fit to all the local potentials considered. The results
plotted as $\times$'s are from Ref.~\protect\cite{lehff} which
displays converged results from separable potentials fitted to the
same parametrization of $\alpha$N phase shifts. For this reason, they
cluster more tightly around their least squares fit.}
\label{Fig:ScatterPlot}
\end{figure}

\begin{figure}
\caption{%
The charge radius of \Li6 versus the binding energy $B$ for breakup
into $\alpha$+$n$+$p$: comparison with other calculations.
Variational results labeled ``K$i$'', $i=1,2,3,6$ are taken from
\protect\cite{kuk1} (``K4'' and ``K5'' have a binding evergy less than 3
MeV), the Faddeev equation results of Bang and Gignoux labeled BG are
taken from \protect\cite{BG}, the integral equation results labeled L
are taken from \protect\cite{lehff}, and the result from the
hyperspherical expansion labeled D is from \protect\cite{Danilin1}.
The Hamiltonian of ``K6'' (which corresponds to Kukulin-84 in Table
\protect\ref{Tab:Compar}) is the same as our model ``$a^*_{\rm Kuk}$''
of Table \protect\ref{Tab:Compar} (plotted here as ``$a^*$ (bare)'')
and the approximate coincidence of the results implies that the radii
of \protect\cite{kuk1} plotted here are not charge radii but bare
radii; see text for details. Note that K3 lies exactly on the
experimental datum. Our results of Fig.~\protect\ref{Fig:ScatterPlot}
are not replotted here, but the straight line through our points is
retained.}
\label{Fig:ScatterPlot2}
\write0{In proof nog punt b toevoegen. nomenclatuur a^* aanpassen}
\end{figure}

\begin{figure}
\caption{%
The  charge form factors of \Li6 versus momentum transfer. Model $a$ does
not have the Coulomb force and model $b$ contains the Coulomb force
including isospin breaking. The contribution of the quadrupole form
factor C2 is also shown for these two models.}
\label{Fig:FormFactor}
\end{figure}

\begin{figure}
\caption{%
Theoretical magnetic moments versus theoretical polarizations of the
neutron in polarized \Li6. The experimental moment $\mu = 0.82205$
nuclear magnetons is indicated as the horizontal line.}
\label{Fig:MuPol}
\end{figure}


\begin{table}

\caption{Parameters of the nucleon-nucleon and alpha-nucleon
potentials used. Parameters are in appropriate powers of MeV and fm.}
\label{Tab:PotParm}

   \begin{center}
      \begin{tabular}{cccccc}
   potential & component & $V_1$ & $\mu_1$ & $V_2$ & $\mu_2$ \\
 \hline
          MT & singlet & $-513.968$ & $1.550$ & $1438.720$ & $3.110$ \\
             & triplet & $-626.885$ & $1.550$ & $1438.720$ & $3.110$ \\
 \hline
         SBB &         & $-47.32$   & $2.30$  & $5.8555$   & $2.30$ \\
 $\rm SBB_2$ & $S$, $P$& $-47.32$   & $2.30$  & $5.8555$   & $2.30$ \\
             & $D$     & $-23.00$   & $2.30$ \\
 $\rm SBB_3$ & $S$     & $+50.00$   & $2.30$ \\
             & $P$     & $-47.32$   & $2.30$  & $5.8555$   & $2.30$ \\
             & $D$     & $-23.00$   & $2.30$ \\
  \hline
         SAT & central  & $-43.0$   & $2.00$  &            & $0.70$ \\
& $\vec{l}\cdot\vec{s}$ & $+40.0$   & $1.50$  &            & $0.35$ \\
      \end{tabular}
   \end{center}
\end{table}

\mediumtext
\begin{table}

\caption{Influence of forbidden-state admixture for model {\em a},
$10\times10$ grid, $j_{\rm max}=2.5$. The number in parentheses is the
estimated uncertainty in the last digit.}
\label{Tab:GammaConv}

   \begin{center}
      \begin{tabular}{ccclccc}
  $\Gamma\,(\hbar^2\rm/M)$ & $E_F\rm\,(MeV)$ & $\braket<H>\rm\,(MeV)$ &
 \multicolumn{1}{c}{$P_F$}
  & $\braket<r^2>^{1/2}\rm\,(fm)$ & $Q\,(e\rm\,fm^2)$ & $P_n^+\rm\,(\%)$ \\
         \hline
  $10^0$ & $-3.998062$ & $-4.8716(4)$ & $5.3750\cdot10^{-3}$ & $2.512(1)$
         & $0.5541(2)$ & $93.1589(1)$ \\
  $10^1$ & $-3.780032$ & $-4.4108(4)$ & $8.3528\cdot10^{-5}$ & $2.565(1)$
         & $0.5612(3)$ & $93.3229(1)$ \\
  $10^2$ & $-3.745151$ & $-4.3373(4)$ & $1.0866\cdot10^{-6}$ & $2.576(1)$
         & $0.5622(3)$ & $93.3309(1)$ \\
  $10^3$ & $-3.740991$ & $-4.3281(4)$ & $1.0637\cdot10^{-8}$ & $2.577(1)$
         & $0.5623(3)$ & $93.3316(1)$ \\
  $10^4$ & $-3.740564$ & $-4.3272(4)$ & $4.9040\cdot10^{-10 }$ & $2.577(1)$
         & $0.5623(3)$ & $93.3316(1)$ \\
  $10^5$ & $-3.740521$ & $-4.3271(4)$ & $4.9909\cdot10^{-10 }$ & $2.577(1)$
         & $0.5623(3)$ & $93.3316(1)$ \\
  $10^6$ & $-3.740517$ & $-4.3270(4)$ & $5.1056\cdot10^{-10 }$ & $2.577(1)$
         & $0.5623(3)$ & $93.3316(1)$ \\
  $10^7$ & $-3.740516$ & $-4.3270(4)$ & $5.1137\cdot10^{-10 }$ & $2.577(1)$
         & $0.5623(3)$ & $93.3316(1)$ \\
$\infty$ & $-3.740516$ & $-4.3270(4)$ & $5.1153\cdot10^{-10 }$ & $2.577(1)$
         & $0.5623(3)$ & $93.3316(1)$ \\
      \end{tabular}
   \end{center}
\end{table}

\begin{table}

\caption{Convergence of the spline approximation for model $a$,
$j_{\rm max}=2.5$.}
\label{Tab:SplineConv}

   \begin{center}
      \begin{tabular}{ccccclc}
   grid & $E_F\rm\,(MeV)$ & $\braket<H>\rm\,(MeV)$ & $E_{\rm Coul}\rm\,(MeV)$
     & $\braket<r^2>^{1/2}\rm\,(fm)$ &
 \multicolumn{1}{c}{$Q\,(e\rm\,fm^2)$} & $P_n^+\rm\,(\%)$ \\
         \hline
 $10\times10$ & $-3.7405(1)$ & $-4.3270(4)$ & $0.8998(2)$
              & $2.577(1)$ & $0.5623(3)$ & $93.3316(1)$ \\
 $14\times14$ & $-3.7840(1)$ & $-4.3375(2)$ & $0.9006(2)$
              & $2.570(1)$ & $0.5482(3)$ & $93.3297(2)$ \\
 $20\times20$ & $-3.7889(1)$ & $-4.3427(2)$ & $0.9011(2)$
              & $2.570(1)$ & $0.55171(7) $ & $93.3267(1)$ \\
 $28\times28$ & $-3.7910(1)$ & $-4.3433(2)$ & $0.9013(2)$
              & $2.570(1)$ & $0.55299(5) $ & $93.3257(1)$ \\
          \hline
 $\infty\times\infty$
              & $-3.7917(4)$ & $-4.3434(2)$ & $0.9014(2)$
              & $2.570(1)$ & $0.5534(2)$ & $93.3254(1)$ \\
      \end{tabular}
   \end{center}
\end{table}

\begin{table}

\caption{Convergence of the partial-wave series for model $a$,
$20\times20$ grid.}
\label{Tab:ChanConv}
   \begin{center}
      \begin{tabular}{cccclll}
  $j_{\rm max}$ & $E_F\rm\,(MeV)$ & $\braket<H>\rm\,(MeV)$
  & $E_{\rm Coul}\rm\, MeV$
  & \multicolumn{1}{c}{$\braket<r^2>^{1/2}\rm\,(fm)$}
  & \multicolumn{1}{c}{$Q\,(e\rm\,fm^2)$}
  & \multicolumn{1}{c}{$P_n^+\rm\,(\%)$} \\
         \hline
         KUK   & $-3.4983(1)$ & $-4.1098(2)$ & $0.8688(2)$ &
 $2.663(1) $ & $0.2232(1)$ & $94.2981(1)$ \\
         $2.5$ & $-3.7889(1)$ & $-4.3427(2)$ & $0.9011(2)$ &
 $2.570(1) $ & $0.55171(7)$ & $93.3267(1)$ \\
         $4.5$ & $-4.2079(1)$ & $-4.4336(2)$ & $0.9304(2)$ &
 $2.4848(8)$ & $0.59616(5)$ & $92.8454(1)$ \\
         $6.5$ & $-4.3655(1)$ & $-4.4481(2)$ & $0.9390(2)$ &
 $2.4595(6) $ & $0.58570(4)$ & $92.7861(1)$ \\
         $8.5$ & $-4.4212(1)$ & $-4.4509(2)$ & $0.9411(2)$ &
 $2.4523(6)$ & $0.57658(4)$ & $92.7876(1)$ \\
 \hline
      $\infty$ & $-4.45(2)$   & $-4.4518(6)$ & $0.9418(5)$ &
 $2.450(2)$  & $0.573(5)$   & $92.788(1)$ \\
      \end{tabular}
   \end{center}
\end{table}

\begin{table}

\caption{Overview of results for thirteen different systems.}
\label{Tab:ModelOverview}

\def\Y{\hphantom{{}^*}}
   \begin{center}
      \begin{tabular}{ccccccc}
  Model & $\braket<H>\rm\,(MeV)$ & $E_{\rm coul}\rm\,(MeV)$
        & $\braket<r^2>^{1/2}\rm\,(fm)$
        & $P_{\alpha\rm D}\,(\%)$ & $Q\,(e\rm\,fm^2)$ & $P_n^+\rm\,(\%)$ \\
         \hline
$a\Y$ & $-4.452(1)$ & $0.942(1)$ & $2.450(2)$ & $67.47(2)$
 & $0.573(5)$ & $92.788(1)$ \\
$b\Y$ & $-3.365(6)$ & $0.932(2)$ & $2.617(4)$ & $68.89(4)$
 & $0.585(5)$ & $92.796(3)$ \\
$c\Y$ & $-3.629(1)$ & $0.892(1)$ & $2.616(1)$ & $70.15(1)$
 & $0.520(1)$ & $93.626(1)$ \\
$d\Y$ & $-3.662(1)$ & $0.869(1)$ & $2.625(1)$ & $71.12(1)$
 & $0.547(1)$ & $93.310(1)$ \\
$e\Y$ & $-4.081(1)$ & $0.923(1)$ & $2.509(6)$ & $67.34(4)$
 & $0.605(4)$ & $92.254(2)$ \\
$f\Y$ & $-3.344(1)$ & $0.927(1)$ & $2.564(9)$ & $71.17(3)$
 & $0.420(2)$ & $94.111(3)$ \\
$g\Y$ & $-4.281(1)$ & $0.934(1)$ & $2.474(4)$ & $66.88(3)$
 & $0.582(3)$ & $93.730(1)$ \\
$h\Y$ & $-4.389(1)$ & $0.952(1)$ & $2.444(2)$ & $62.42(2)$
 & $0.256(1)$ & $98.123(2)$ \\
$a^*$ & $-3.516(2)$ & $0.910(1)$ & $2.542(4)$ & $69.05(4)$
 & $0.578(5)$ & $92.953(2)$ \\
$e^*$ & $-3.162(1)$ & $0.883(1)$ & $2.646(3)$ & $69.46(5)$
 & $0.598(4)$ & $92.559(2)$ \\
$f^*$ & $-2.444(1)$ & $0.841(2)$ & $3.040(8)$ & $75.64(8)$
 & $0.405(8)$ & $94.275(4)$ \\
$g^*$ & $-3.352(1)$ & $0.900(1)$ & $2.585(2)$ & $68.62(3)$
 & $0.589(3)$ & $93.896(1)$ \\
$h^*$ & $-3.440(1)$ & $0.919(1)$ & $2.544(2)$ & $63.98(2)$
 & $0.258(1)$ & $98.277(1)$ \\
      \end{tabular}
   \end{center}
\end{table}

\begin{table}
\caption{Comparisons with other calculations.}
\label{Tab:Compar}

\def\Y{\hphantom{^a}}
\def\0{\hphantom{0}}
      \begin{tabular}{cccccccc}
  Model & $\braket<H>\rm\,(MeV)$ &  $\braket<r^2>^{1/2}\rm\,(fm)$ &
 $Q\,(e\rm\,fm^2)$ & $P_S\,(\%)$ & $P_P\,(\%)$ & $P_{P'}\,(\%)$ & $P_D\,(\%)$\\
         \hline
 $a^*_{\rm Kuk}$
           & $-3.29$ & $2.39$%
\tablenote{%
This is the {\em bare} radius, {\em i.e.}, without taking into account
the electromagnetic size of the constituents. We believe that it is
this radius which should be compared to that labeled Kukulin-84. The
full radius ({\em i.e.}, the radius one gets if the electromagnetic
size of the constituents is taken into account) is $2.79\rm\,fm$. Note
that some authors give matter radii instead on charge radii. These two
quantities are identical if the $T=1$ component of the wave function
is ignored, and the constituent charge radii are taken to be equal to
the constituent matter radii.}
                                & $+0.21$ & $91.43$ & $0$ & $2.35$ & $6.22$\\
 Kukulin-84 \cite{kuk1}
           & $-3.39$ & $2.35\Y$ & $+0.24$ & $91.54$ & $0$ & $0.27$ & $8.19$\\
 Kukulin-86 \cite{kuk2}
           & $-3.85$ &          &         & $95.54$ & $0$ & $1.08$ & $3.38$\\
  Kukulin-92 \cite{kuk92}
           & $-3.33$ & $2.52\Y$  & $+0.40$ & $90.16$ &
                                                    $0.15$ & $2.12$ & $7.55$\\
 $b$       & $-3.37$ & $2.62\Y$  & $+0.59$ & $89.48$ &
                                                    $0.22$ & $2.65$ & $7.65$\\
         \hline
 Bang \cite{BG}
         & $-3.20$ & $2.44\Y$ & pos. & $92.8$ & $0.7\0$ & $5.3\0$ & $1.2\0$\\
         \hline
 Danilin \cite{Danilin1}
       & $\sim-3.4$ & $2.48\Y$ & $+0.40$ & $93.04$ & $0.24$ & $3.34$ & $3.38$\\
         \hline
proj. (4\%) \cite{lehff,lehmm}
           & $-3.90$  & $2.43\Y$ &       & $91.47$ & $0.48$ & $4.65$ & $3.40$\\
rep. (4\%) \cite{lehff,lehmm}
           & $-4.06$  & $2.40\Y$ &       & $91.78$ & $0.50$ & $4.01$ & $3.71$\\
      \end{tabular}
\bigskip

\end{table}

\begin{table}

\caption{Probabilities in eight different systems.}
\label{Tab:ProbsOverview}

\def\Y{\hphantom{^*}}
   \begin{center}
      \begin{tabular}{ccccccc}
  Model & $P_S\,(\%)$ & $P_{P}\,(\%)$ & $P_{P'}\,(\%)$ & $P_D\,(\%)$
        & $P_n^+[/P_p^+]\rm\,(\%)$ & $\mu (\mu_N)$ \\
         \hline
 $a\Y$ & $89.290(2)$ & $0.232(2)$ & $2.808(4)$ & $7.667(2)$
       & $92.788(1)$ & $0.82502(3)$ \\
 $a^*$ & $89.590(1)$ & $0.647(1)$ & $2.166(4)$ & $7.596(2)$
       & $93.057(1)$ & $0.82708(3)$ \\
 $b\Y$ & $89.478(2)$ & $0.223(4)$ & $2.647(3)$ & $7.650(3)$
       & $92.796(3)/92.982(2)$ & $0.82575(3)$ \\
 $c\Y$ & $90.626(1)$ & $0.216(1)$ & $2.861(1)$ & $6.519(1)$
       & $93.626(1)$ & $0.83139(2)$ \\
 $d\Y$ & $90.163(1)$ & $0.206(1)$ & $2.341(1)$ & $7.291(1)$
       & $93.310(1)$ & $0.82899(2)$ \\
 $e^*$ & $88.712(2)$ & $0.441(1)$ & $3.218(1)$ & $7.630(3)$
       & $92.559(2)$ & $0.82328(2)$ \\
 $f^*$ & $92.119(5)$ & $0.105(1)$ & $0.536(1)$ & $7.239(3)$
       & $94.275(4)$ & $0.83633(2)$ \\
 $g^*$ & $90.881(1)$ & $0.237(1)$ & $2.463(3)$ & $6.418(1)$
       & $93.896(1)$ & $0.83344(2)$ \\
 $h^*$ & $96.573(1)$ & $0.197(1)$ & $2.995(1)$ & $0.236(1)$
       & $98.277(1)$ & $0.86671(2)$ \\
      \end{tabular}
   \end{center}
\end{table}


\begin{references}
\bibitem{SAID}
The current NN phase shifts can be obtained interactively through the
Scattering Analysis Interactive Dial-in (SAID) program at VPI.
\end{references}
\end{document}